\documentclass[fleqn,12pt,twoside]{article}
\usepackage{espcrc1,multirow}
\readRCS
$Id: espcrc1.tex,v 1.2 2004/02/24 11:22:11 spepping Exp $
\usepackage{graphicx}
\usepackage[figuresright]{rotating}
\usepackage{amssymb}

\newcommand{\AmS}{{\protect\the\textfont2
  A\kern-.1667em\lower.5ex\hbox{M}\kern-.125emS}}

\begin{document}

\title{\bf Very high-energy observations of the two high-frequency \\
peaked BL Lac objects 1ES 1218+304 and H~1426+428 \\}

\author{
C. Mueller\address[McGill]{Department of Physics, McGill University \\
Montreal, QC H3A 2T8, Canada}$^1$\footnotetext{$^1$present address: Sander Geophysics Limited, 
Ottawa, ON  K1V 1C1},
N. Akhter\address[Barnard]{Department of Physics and Astronomy, Barnard College, 
Columbia University \\ 
New York, NY 10027, USA},
J. Ball\address[UCLA]{Department of Physics and Astronomy, University of California, 
Los Angeles \\ Los Angeles, CA 90095, USA}$^2$\footnotetext{$^2$present address: Gemini North, Hilo, HI 96720, USA},
D.A. Bramel\addressmark[Barnard]$^3$\footnotetext{$^3$present address: Interactive Brokers, Greenwich, CT 06830, USA},
J. Carson\addressmark[UCLA]$^4$\footnotetext{$^4$present address: The Claremont Colleges, Joint Sciences 
Department, Claremont, CA 91711-5916, USA},
C.E. Covault\address
[Case]{Department of Physics, Case Western Reserve University\\ 
Cleveland, OH 44106, USA},
D. Driscoll\addressmark[Case]$^5$\footnotetext{$^5$present address: Kent State University, Ashtabula, OH 44004, USA},
P. Fortin\addressmark[Barnard]$^6$\footnotetext{$^6$present address: Laboratoire Leprince-Ringuet, 
\'Ecole Polytechnique, CNRS/IN2P3, Palaiseau, France},
D.M. Gingrich\address[Alberta]
{Centre for Particle Physics, Department of Physics, University of Alberta\\
Edmonton, AB T6G 2G7, and \\
TRIUMF, Vancouver, BC V6T 2A3, Canada},
D.S. Hanna\addressmark[McGill], 
A. Jarvis\addressmark[UCLA]$^7$\footnotetext{$^7$present address: Disney Interactive Media Group, North
Hollywood, CA 91601, USA},
J. Kildea\addressmark[McGill]$^8$\footnotetext{$^8$present address: McGill University, Medical Physics Unit, 
Montreal, QC H3G 1A4},
T. Lindner\addressmark[McGill]$^9$\footnotetext{$^9$present address: Department of Physics
and Astronomy, University of British Columbia, Vancouver, BC  V6T 1Z1, Canada},
R. Mukherjee\addressmark[Barnard],
R.A. Ong\addressmark[UCLA], 
K. Ragan\addressmark[McGill],
R.A. Scalzo\address[Chicago]{Department of Physics, University of Chicago \\
Chicago, Il 60637, USA}$^{10}$\footnotetext{$^{10}$present address: Department of Physics, Yale University, New Haven, CT 06520, USA},
D.A. Williams\address[UCSC]
{Santa Cruz Institute for Particle Physics, University of California, Santa Cruz \\ 
Santa Cruz, CA 95064, USA},
J. Zweerink\addressmark[UCLA]
}

\maketitle

\newpage 

\begin{abstract}

\noindent
{\bf Abstract}

We present results of very-high-energy gamma-ray observations ($E_\gamma > 160$ GeV) of 
two high-frequency-peaked BL Lac (HBL) objects, 1ES 1218+304 and H~1426+428, 
with
the Solar Tower Atmospheric Cherenkov Effect Experiment (STACEE). 
Both sources are very-high-energy gamma-ray emitters above
100 GeV, detected using ground-based Cherenkov telescopes. 
STACEE observations of 1ES 1218+304 and H~1426+428 did not produce
detections; we present 99\% CL flux upper limits for both sources,
assuming spectral indices measured mostly at higher energies. 

\end{abstract}

\section{Introduction}

Nearly thirty active galaxies of the ``blazar'' class have been detected as 
very-high-energy gamma-ray sources using ground-based atmospheric Cherenkov
telescopes (ACTs) (e.g. see review \cite{Hinton2009}). Blazars are compact,
highly variable extragalactic objects, characterized by 
non-thermal continuum emission that extends from radio to very-high-energy
gamma rays. These sources are broadly classified into two groups: BL 
Lacertae (BL Lac) objects and 
flat-spectrum radio quasars (FSRQs). BL Lac objects 
generally have smooth, featureless continuum spectra, with emission
lines that are weak or absent. 
The spectral energy distributions (SEDs) of these sources 
typically have two broad peaks, one at low energies (radio
to X-ray) and the other at higher energies (keV to TeV). 
Blazars are thought to be 
highly beamed sources, with relativistic jets oriented close to
the line of sight \cite{Urry1995}. 
In the current blazar paradigm, 
the low energy peak in the blazar SED is explained as synchrotron emission from 
high-energy electrons in the jet, while the high-energy emission is due to
relativistic charged particles in the blazar jet. In
leptonic models, high-energy gamma rays are produced by
inverse Compton (IC) scattering of ambient low energy photons by
relativistic electrons. In an alternate scenario, 
hadronic models explain the gamma-ray energy emission as due to 
neutral pions produced by 
energetic protons (e.g. see
\cite{Boettcher2007} and \cite{Muckeetal2003} for reviews). 

Blazars are broadly categorized into sub-groups based on the synchrotron peak
frequencies and the relative power in the low and high-energy peaks of
their SEDs \cite{Fossati1998}. 
Of the blazars belonging to the BL Lac class, 
low-frequency-peaked blazars (LBLs) have this peak in the radio or 
optical band while for high-frequency-peaked blazars (HBLs), it is in the X-ray band. 
Historically, the majority of the EGRET-detected
blazars belong to the FSRQ class, with the synchrotron peak in the
radio-optical band \cite{Hartman1999}. 
In its first 5.5 months of observations, {\it Fermi} detected 21 TeV-selected blazars 
of which 13 are HBLs \cite{Fegan2009}.
Since then, the list of Fermi-detected blazars has grown and the first catalog of active 
galactic nuclei (AGN) detected using the {\it Fermi} Large Area Telescope includes 
671 gamma-ray sources
located at high galactic latitudes that are associated statistically with AGNs \cite{Abdo2010}. 
All but a handful of TeV
blazars detected to date belong to the HBL category. The
exceptions include 
BL Lacertae, detected using the MAGIC
telescope \cite{Albert2006}, and W Comae, 3C 66A, and PKS1424+240, recently 
detected in TeV gamma rays by the VERITAS (\cite{Acciari2008_WComae},
\cite{Acciari2009},\cite{Acciari2009b}) collaboration. 
STACEE carried out an extensive observing campaign on two LBLs, 3C 66A and OJ 287, but
did not detect any significant gamma-ray emission from either source 
(\cite{Lindner2007}, \cite{Bramel2005}).
The only blazar detected using STACEE was the HBL Mrk 421 in observations
carried out in 2001 \cite{Boone2002} and 2004, when STACEE made the first 
measurement of the differential energy spectrum of the source between
130 GeV and 400 GeV \cite{Carson2007}. 

HBLs have been predicted to be good candidates for TeV gamma-ray
emission, based on synchrotron self-Compton (SSC) emission models
\cite{Costamante2002} as well as hadronic models
\cite{Mannheim1993}. Several of the ``extreme'' synchrotron BL Lacs
\cite{Costamante2001} have been detected at TeV energies, confirming
these predictions. Both H~1426+428 and 1ES 1218+304 were predicted to
be TeV gamma-ray emitters, and this was part of the motivation for
STACEE to observe these sources. 
Neither of these sources was detected by EGRET at GeV energies, while both 
have been detected with {\it Fermi}. 

1ES 1218+304 is an X-ray-bright (flux in the 2-10 keV range $\sim$ 2$\times 10^{-11}$
erg cm$^{-2}$ s$^{-1}$\cite{Sato2008}) HBL, and
at a redshift of $z = 0.182$, it is one of the more distant VHE blazars detected
to date. A detection of 1ES 1218+304 was recently reported by both MAGIC 
\cite{Albert2006} and VERITAS \cite{Fortin2008}, at
energies $>100$ GeV, providing further evidence that X-ray-bright HBLs tend
to be strong VHE sources. The source is detected with {\it Fermi} with no
evidence for variability \cite{Fegan2009}.
The MAGIC detection of TeV emission from this source motivated the observations 
by STACEE that were carried out in the 2006 and 2007 observing seasons. 

H~1426+428 is classified as an ``extreme'' BL Lac, with its
synchrotron peak at an energy greater than 100 keV; it has long been 
predicted to be a TeV emitter. The source was
first detected at TeV energies by the Whipple collaboration \cite{Horan2002} 
and later
confirmed using other ground-based imaging atmospheric Cherenkov telescopes 
\cite{Aharonian2002,Djannati-Atai2002}. Like 1ES 1218+304, its
distance ($z = 0.129$) makes it a promising candidate for studying
the extragalactic infra-red background radiation. 
{\it Fermi} has detected weak emission from H~1426+428 at 
energies less than 20 GeV \cite{Fegan2009}. 
STACEE carried out observations of
H~1426+428 in 2003 and 2004. 

The Solar Tower Atmospheric Cherenkov Effect Experiment (STACEE) was a
ground-based experiment sensitive to gamma rays above 100
GeV. STACEE operated from 2001 until its de-commissioning in the summer of 2007
\cite{Gingrich2005}. 
STACEE observed several active galaxies of the blazar class with
the aim of understanding particle acceleration and emission mechanisms
in these sources. With an energy threshold close to 100 GeV, STACEE
also had the potential to study the effect of the extragalactic
background light (EBL) on the spectra of distant blazars. 
\S2 gives a brief overview of the
STACEE instrument. In \S3 we describe the data taking and observing 
strategy of STACEE, and in \S4 we present results from STACEE observations of 
1ES 1218+304 and H~1426+428.

\section{The STACEE Detector}

The STACEE detector used the atmospheric Cherenkov technique, reconstructing 
very-high-energy gamma rays by observing Cherenkov photons resulting from
the gamma-ray interactions in the upper atmosphere. The instrument
used 64 of the 212 heliostats of the National Solar Thermal Test
Facility (NSTTF), near Albuquerque, NM, to direct the Cherenkov light to
secondary mirrors, and then onto cameras composed of photomultiplier
tubes (PMTs), such that each PMT viewed only a single
heliostat. 
Each heliostat had an
area of 37 ${\rm m}^2$, leading to a large primary mirror area of
approximately 2400 ${\rm m}^2$ and sensitivity to low photon densities
(ie, to low gamma-ray energies). The recorded PMT information was used
as a measure of the Cherenkov photon impact points on the ground;
analysis of the time of arrival of the Cherenkov wavefront allowed
the reconstruction of the arrival direction and energy of the gamma ray, and
it improved the rejection of charged cosmic rays which constituted the
main background.  

The Cherenkov wavefront is typically only a few nanoseconds in 
duration, and STACEE used three-level trigger electronics to record
Cherenkov events. Each PMT signal was AC-coupled and then split, with
one copy going to a rapid (1 GS/s)
8-bit digitizer and one copy going to a discriminator set to detect
pulses larger than approximately 5 photoelectrons. The discriminated
logic signals were then sent to a custom trigger system
\cite{Gingrich2005,Martin2000} which examined the 64 signals in eight clusters of
eight channels. Each cluster was considered to trigger if at least five of
its constituent channels contained pulses in a 24-nanosecond window, and
an event trigger was generated if five of the eight clusters triggered in
a 16-nanosecond window. More details of the STACEE detector and operations 
are given elsewhere \cite{Gingrich2005,Hanna2002}. 


\section{Data Taking and Observing Strategy} 

STACEE recorded astrophysical source data
in an ``ON-OFF'' observing mode consisting of a 28-minute
ON-source run where the source was tracked at the centre of the field
of view, followed by an OFF-source run of the same duration, when a
patch of the sky at identical declination but 30 minutes ahead or
behind the source in right ascension was observed. The ``OFF'' run was
used to determine the background, based on the facts that a gamma-ray
excess is expected only from the source direction, and that the rate of
background showers from charged cosmic rays is the same for both
runs. The gamma-ray flux was inferred from the difference in count
rates between the ON-OFF pair of runs for a particular source
observation. Prior to astrophysical source observations, the discriminator
threshold was adjusted 
to eliminate noise triggers resulting from random
coincidences of the night sky background. The typical STACEE trigger rate 
was about 5 Hz. Further details of STACEE
nightly operations are given elsewhere \cite{Lindner2007}.

The STACEE data set was analyzed to remove data taken in unfavorable
weather conditions or with detector malfunctions (e.g. malfunctioning 
heliostats, high voltage trips, etc.), in order to eliminate biases in the
trigger rates and to increase the sensitivity of the instrument. The STACEE
data cleaning criteria are described elsewhere
\cite{Bramel2005}. In addition, field brightness corrections 
had to be applied to account for the differences between the relative
brightness of the ON and OFF fields using a technique called {\sl 
padding} \cite{Scalzo2004}\footnote{Typical field brightness differences led to 
a difference of trigger rate between ON and OFF fields of a few percent}. 
A crucial step in the STACEE data
analysis is the rejection of Cherenkov showers generated by the
charged particle background. 
STACEE used the {\sl grid alignment}
technique for cosmic ray background
suppression \cite{Smith2006,Kildea2005a,Lindner2007}; this technique uses
the differences in the distribution of Cherenkov light on the ground for gamma ray and cosmic ray
induced showers to provide gamma-hadron separation. Using this analysis
method, STACEE reported a detection of the Crab Nebula at a
significance of $8.1\sigma$ in a data set of 21 hours of observation in 2002-2004 
\cite{Kildea2005b}. 

\section{Results from STACEE Observations of 1ES 1218+304 \& H~1426+428}  

\subsection{1ES 1218+304} 
\bigskip

STACEE observed 1ES 1218+304 during the 2006 and 2007 observing
seasons for a total of 70.9 hours (152 ON-OFF pairs) \cite{Mukherjee2007}. This data set was
reduced to 28.3 hours after the standard data quality cuts (referred to
above) were applied to the data. The differences in the field 
brightnesses between the ON and the 
OFF fields was also taken into account. Table~1 summarizes the livetime available 
for the 2006 and 2007 data sets. A net 
ON-source excess of 236 events was seen, compared to a background of 5547
events, corresponding to a statistical significance of $2.3\sigma$ (calculated using 
equation (17) in \cite{LiMa1983}). In the 152 ON-OFF pairs,
there are no individual significances above $4\sigma$. 
The $2.3\sigma$ excess is not statistically significant, and we choose to
calculate a flux upper limit for the source, as described
below. 



\begin{table}[b!]

\caption{Summary of STACEE data on 1ES 1218+304.}
\begin{center}
\vskip 0.05in
\begin{tabular}{llrrrc}\hline\hline

Year      &  Livetime & ON-source & OFF-Source  & Excess    & Significance\\
          &  (hrs)    & Events    & Events      & Events    & 		  \\
\hline\hline

2006 & 12.0 & \multirow{2}{*}{5547} & \multirow{2}{*}{5311} & \multirow{2}{*}{236} & \multirow{2}{*} {2.3$\sigma$}\\

2007 & 16.3 &  & & &  \\
       	                 
\hline

\end{tabular}
\end{center}

\end{table}

\begin{figure}
\begin{center}
\includegraphics*[width=0.8\textwidth, angle=90,height=3in]{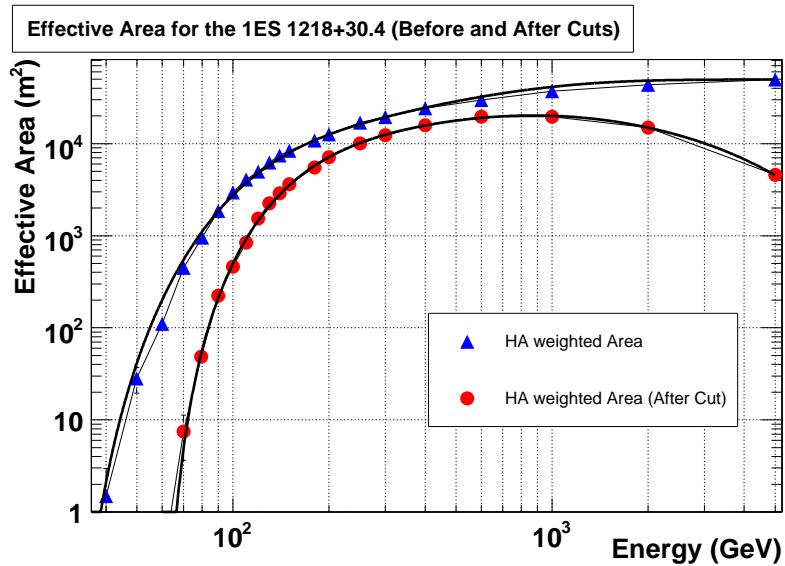}
\end{center}
\caption{\label{effarea}Average effective area of the STACEE detector for 
  the 1ES 1218+304 observations, as a function of gamma-ray energy, after 
  weighting in hour-angle by the time spent at each pointing. The effective
  area after background rejection cuts is also shown. The solid lines are meant to
  guide the eye only.}
\end{figure}

In order to interpret the 1ES 1218+304 data and calculate flux upper
limits, extensive Monte Carlo simulations of the STACEE detector had
to be carried out to determine the effective area of the instrument
(see, for example, \cite{Bramel2005}) as a function of energy and arrival
directions for both gamma ray and cosmic ray showers. 
Figure~\ref{effarea} shows the
hour-angle-averaged effective area of STACEE for the 1ES 1218+304
observations, weighted according to how much time was spent at each
1ES 1218+304 pointing. The effective area after background rejection cuts 
is also shown in the figure. 
As described in \cite{Lindner2007}, the effective area
of STACEE below 100 GeV is reduced as a result of the cuts, partly due to the increased
pulse-height threshold that is part of the padding process. At
energies above $\sim$ 1 TeV, the effective area after cuts is lowered due to
the grid alignment cut applied to the data. 

The effective area for the 1ES 1218+304 data was used to
calculate the detector energy threshold and flux upper limits. We
assumed a power-law differential energy spectrum 
${{dN}\over{dE}} \sim E^{-\Gamma}$ with a spectral index of $\Gamma=3.0$ 
for 1ES 1218+304, based on the
MAGIC \cite{Albert2006} and VERITAS \cite{Fortin2008} measurements. 
Folding this with the effective area
curve, we get an energy threshold $E_{th}$ of 
$\sim 155\pm 28_{\rm sys}$ GeV (where the energy threshold is defined by 
convention as the peak of the resulting detector response function). 
The systematic uncertainty of 28 GeV
arises primarily from uncertainties on the optical characteristics of the experiment
(atmospheric attenuation, optical alignment, mirror reflectivities). 

\begin{figure}
\begin{center}
\includegraphics*[width=0.8\textwidth,height=3.0in]{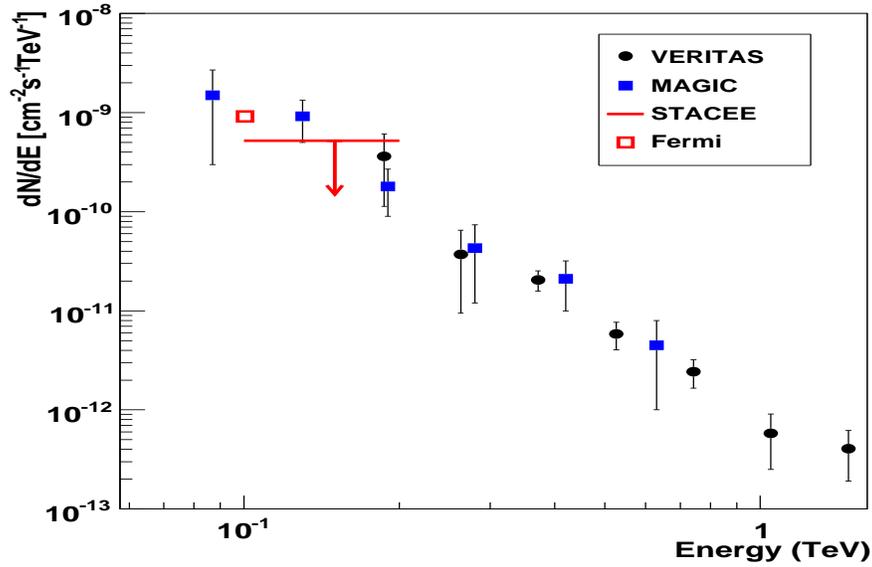}
\end{center}
\caption{Gamma-ray spectrum of 1ES 1218+304, as measured by VERITAS
  \cite{Fortin2008} and MAGIC \cite{Albert2006}, with the 
  STACEE 99\% flux
  upper limit at 155 GeV. 
  The {\it Fermi} point is an extrapolation from lower energy data to 100 GeV.
  Note that the different measurements are not contemporaneous. } \label{fig4}
\end{figure}

We derive a 99\% confidence level (CL) flux upper limit at 155 GeV of
$4.7\times 10^{-10}$ cm$^{-2}$ s$^{-1}$ TeV$^{-1}$. 
Figure~\ref{fig4} shows the STACEE upper limit
overlaid on the VERITAS and MAGIC spectra. 
The {\it Fermi} point is an extrapolation to 100 GeV
from measurements at lower energies, using the {\it Fermi} spectral index and no EBL model. 
The STACEE measurements were carried out at different 
epochs than the VERITAS, MAGIC, and {\it Fermi} measurements. 
Therefore, the STACEE upper limit is not necessarily in
conflict with the two lowest-energy MAGIC data points and the {\it Fermi} extrapolation. 
Blazars are known to be variable sources, and it is possible that the source was not active
when STACEE observed it.  

\subsection{H~1426+428}
\bigskip

STACEE observations of the HBL H~1426+428 were performed in 2003 and 2004. 
The detector was 
configured differently for the two data sets. 
In the 2003 data, the heliostats were canted 
(tilted) to receive light from the position where the air shower 
created by a primary VHE gamma ray 
contains a maximum number of charged 
particle secondaries, 
at an altitude of approximately 12 km. 
The canting scheme was changed for the 2004 observations 
to allow for several heliostats to be aimed directly at 
the source being observed ({\it i.e.} these heliostats were not canted). 
This change allowed for a better reconstruction of low-energy
events and improved background rejection. 
Because of the different 
detector sensitivities, the data sets must be treated independently and each one 
compared to appropriate Monte Carlo simulations. 
A total of 52.5 hours of data were recorded, with 28.5 hours remaining after standard data 
quality cuts.  
Table~2 summarizes the data sets, including the number of ON-source and OFF-source events 
after selection cuts, and the significance of the ON-source excess.  

\begin{table}
\caption{Summary of STACEE data on H~1426+428.}

\begin{center}
\vskip 0.05in
\begin{tabular}{llrrrc}\hline\hline

Year      &  Livetime & ON-Source     & OFF-source & Excess & Significance \\
          &  (hrs) & Events        & Events     & Events &   \\
\hline\hline

2003 & 8.9 & 24480 & 24162 & 318 & 1.6$\sigma$\\

2004 & 19.7 & 49093 & 49065 & 28 & 0.2$\sigma$ \\
       	                 
\hline

\end{tabular}
\end{center}
\end{table}

As was done for the 1ES~1218+304 data, 
detailed Monte Carlo simulations of the detector were
used to calculate the effective area. The simulations were done separately for the
two canting schemes and for several detector pointing directions (hour-angles).
Weighting according to the hour-angle
distribution of the data then results in an overall effective area 
function (for each data set); these curves are similar 
to those of Figure~\ref{effarea}. 

Each effective area curve is then folded with a power law spectrum to generate 
the detector response function. For H~1426, we assume a spectral index 
of $-3.50$ (\cite{Horan2002},\cite{Petry2002},\cite{Djannati-Atai2002}).
The resulting STACEE response function peaks at 
164 GeV; the estimated systematic uncertainty on this value is 28 GeV. 

\begin{figure}
\begin{center}
\includegraphics*[width=0.8\textwidth,height=3.0in]{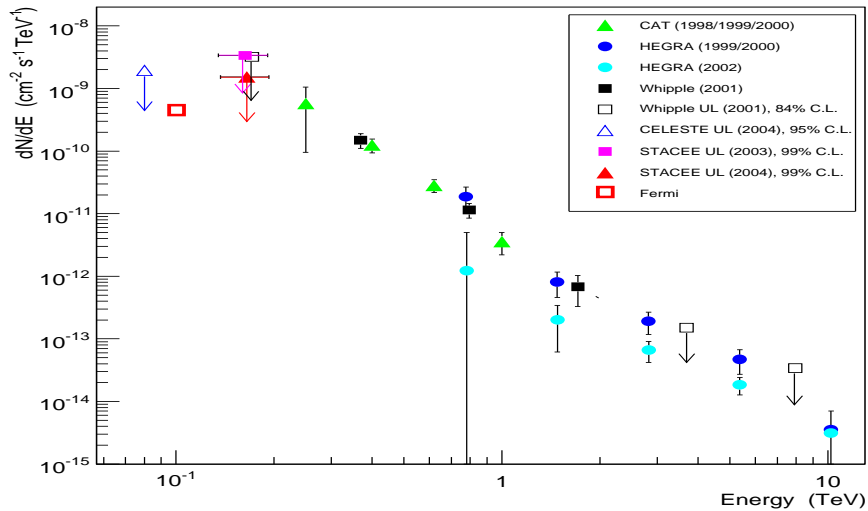}
\end{center}
\caption{Gamma-ray spectrum of H 1426+428 with the 
  STACEE 99\% flux upper limits at 164 GeV. Most other measurements and 
  upper limits are from imaging Cherenkov telescopes.
  The {\it Fermi} point is an extrapolation from lower energy data to 100 GeV.
  Note that the measurements are not contemporaneous. } \label{fig5}
\end{figure}

The detector response functions and the measured rate of excess events are 
then used to derive 99\% CL flux upper limits at E$_{th}$ of 164 GeV of
$3.4\times 10^{-9}$ cm$^{-2}$ s$^{-1}$ TeV$^{-1}$ 
and 
$1.5\times 10^{-9}$ cm$^{-2}$ s$^{-1}$ TeV$^{-1}$ 
for the 2003 and 2004 data sets, respectively. 
These results are shown in Figure~\ref{fig5} together with
measurements at 
higher energies by imaging Cherenkov telescopes, and an extrapolation of lower energy 
{\it Fermi} data to our energy range (again using the measured {\it Fermi} spectral index and no
EBL model). 
 
\bigskip

\section{Conclusions and Summary}

Both 1ES 1218+304 and H 1426+428 are X-ray bright HBLs, and were promising targets 
for STACEE based on their broad-band SEDs. The main motivation for the STACEE 
observations was to obtain data on these two blazars in a largely-unexplored 
energy range. 
Together with observations of H 1426+428 made with the 
CELESTE instrument \cite{Smith2006}, 
the data presented in this paper represent the only observations of these blazars 
below 200 GeV by solar heliostat arrays. 

Both sources have now been detected using {\it Fermi}'s LAT instrument, with 
hard spectra \cite{Fegan2009}. Indeed, their spectra rank as among the hardest 
of the 38 TeV-selected AGN that {\it Fermi} has observed, reaffirming the interest of
having observational data in the STACEE energy range.  

In the case of 1ES 1218+304, the extrapolation of the measured {\it Fermi} GeV 
spectrum to the STACEE threshold energy $E_{th}$ results in a differential flux that is less
than a factor of two below the upper limit reported here. Our upper limit is also comparable
to the measured MAGIC flux in this energy range, and to extrapolations
of higher-energy observations by VERITAS (see Figure 2). 

In the case of H1426+428, the {\it Fermi} GeV spectrum is nearly an order of magnitude below
our upper limit reported here. However, our limit is comparable to the extrapolation from 
higher energies of imaging telescope data (see Figure 3). This again points out the importance
of data in the energy range of the solar-array telescopes like STACEE - this is the range where
the GeV spectrum ({\it Fermi} measures a spectral index of $\Gamma\approx 1.5$ \cite{Fegan2009})
transitions to the TeV spectrum (with a spectral index of 3.5).

Many AGNs are known to be highly variable sources in the VHE regime. For example, VERITAS 
recently observed a flare from 1ES 1218+304 with an estimated flux-doubling time of one day 
\cite{Imran2010}. For both sources reported here, the STACEE data was accumulated over several years, 
and the upper limits represent mean flux levels over the observational period.
The energy range reported on here is worthy of investigation by higher sensitivity instruments such 
as a new generation of imaging arrays. 

\bigskip

In summary, we have presented data from STACEE observations of two 
HBL candidates suggested as potential TeV emitters by Costamante and 
Ghisellini \cite{Costamante2002}, 1ES 1218+304 and H~1426+428. 
We have not detected a signal from either of these sources and have set upper
limits on their gamma-ray flux levels, at an energy below that of most of the 
imaging atmospheric Cherenkov telescopes detections.

\section{Acknowledgements}

We are grateful to the staff at the National Thermal Solar Test Facility 
for their enthusiastic and professional support. 
This work was funded in part by the U.S. National Science Foundation, 
the University of California, Los Angeles,
the Natural 
Sciences and Engineering Research Council, le Fonds Qu\'ebecois de la 
Recherche sur la Nature et les Technologies,
the Research Corporation, and the 
California Space Institute.

\end{document}